\documentclass{aastex6}

\begin{document}
\bibliographystyle{plainnat}


\title{Experimental phase functions of mm-sized cosmic dust grains}



\author{O. Mu\~{n}oz\altaffilmark{1}, F. Moreno\altaffilmark{1},
 F. Vargas-Mart\'{i}n \altaffilmark{2}, D. Guirado\altaffilmark{1},  J. Escobar-Cerezo\altaffilmark{1}, M. Min\altaffilmark{3,4}, J.W. Hovenier\altaffilmark{4}}
\affil{{1} Instituto de Astrof\'{i}sica de Andaluc\'{i}a, CSIC, 
Glorieta de la Astronom\'{i}a s/n, 18008 Granada, Spain.} 
\affil{{2} Department of Electromagnetism and Electronics,  University of Murcia, 30100 Murcia, Spain.} 
\affil{{3} SRON Netherlands Institute for Space Research, Sobornnelaan 2, 3584 CA Utrecht, The Netherlands.} \and
\affil{{4} Astronomical Institute ``Anton Pannekoek'', University of Amsterdam Science Park 904 1098 XH, Amsterdam, The Netherlands.}

\begin{abstract}
We present experimental phase functions of three types of millimeter-sized dust grains consisting of enstatite, quartz and volcanic material from Mount Etna, respectively. The three grains present similar sizes but different absorbing properties. The measurements are performed at 527 nm covering the scattering angle range from 3 to 170 degrees. The measured phase functions show two well defined regions i) soft forward peaks and ii) a continuous increase with the scattering angle at side- and back-scattering regions. This behavior at side- and back-scattering regions are in agreement with the observed phase functions for the Fomalhaut and HR 4796A dust rings. Further computations and measurements (including polarization) for millimeter sized-grains are needed to draw some conclusions about the fluffy or compact structure of the dust grains.     

\end{abstract}




\section{Introduction} 
\label{sec:intro}

Small dust particles are ubiquitous in many different astrophysical bodies ranging from planetary and cometary atmospheres,
 zodiacal clouds, planetary rings to  debris disks and protoplanetary systems. The way those small particles scatter and
 absorb stellar light affects the thermal structure of the body under study and subsequently its chemical and dynamical properties. The spectral dependence of light scattered by those particles is widely used for retrieving the physical properties of the grains and their spatial distribution.

Dust grains in planetary
 atmospheres (\citet{rages1983}, \citet{tomasko1999}, \citet{wolff2010}), debris and protoplanetary disks (\citet{weinberger1999}) usually produce strong forward scattering and nearly a flat dependence of the scattering angle at side- and back-scattering regions.  This seems to indicate the presence of compact and/or aggregate 
dust grains with sizes ranging from sub-micron up to tens of microns. Those findings are in agreement with laboratory measurements of cosmic dust analogues (\citet{volten2006}, \citet{volten2007}, \citet{laan2009}, \citet{munoz2012}, \citet{dabrowska2015}).   

There are some astronomical observations that indicate the presence of millimeter-sized cosmic dust grains. Such is the case of comet 
67P/Churyumov-Gerasimenko, target of the ESA Rosetta mission. Main conclusions from the GIADA (Grain Impact Analyser and Dust Accumulator) instrument provide a scenario in which we can find compact particles with sizes ranging from  0.03 mm up to 1 mm and fluffy aggregates from sub-micron up to 2.5 mm (\citet{fulle2015}).  Moreover, pre-perihelion observations of the OSIRIS (Optical, Spectroscopic, and Infrared Remote Imaging System) cameras indicate that dust optical scattering is dominated by 100 
$\mu$m to millimeter-sized grains (\citet{rotundi2015}).

 Apart from comets, circumstellar disks also can host large cosmic dust grains (e.g. \citet{andrews2005},  \citet{canovas2015}, \citet{kataoka2016}, \citet{canovas2016}). Interesting case studies related to circumstelar disks are reported by \citep{kalas2005} and \citet{milli2017} for dust orbiting Fomalhaut and HR 4796A, respectively. HST imaging of Fomalhaut shows that a very small fraction of the stellar light is scattered into our line of sight. Moreover, the phase function at side- and back-scattering regions increases with the scattering angle  \citep{lebouquin2009}. A similar behavior has been recently reported by \citet{milli2017} for the HR 4796A dust ring.
This scattering behavior could be caused by the presence of large grains; r $\geq$ 100 $\mu$m  in the Fomalhaut, and r$\sim$ 30 $\mu$m in the HR 4796 A dust rings, respectively (\citet{min2010}, \citet{milli2017}).  
Subsequent analysis of far-infrared images of Fomalhaut obtained with the Herschell Space Observatory (\citet{acke2012}) indicates that the belt around Fomalhaut could consist of fluffy aggregates. The main reason is that large compact grains do not have the thermal properties needed to explain the far-infrared images. On the contrary, fluffy aggregates consisting of small monomers  could have the absorption properties of small grains showing the scattering anisotropy of large particles. Still, simulated phase functions of fractal aggregates  (e.g. \citet{bertini2007}, \citet{misha2007}, \citet{moreno2007}, \citet{okada2008}, \citet{min2016}) and experimental phase functions of aggregates with sizes larger than the wavelength of the incident light  (\citet{volten2007}), provide a strong diffraction spike but do not reproduce the observed slope of the phase function at side- and back-scattering angles. We might note that numerical scattering simulations of fractal aggregates are limited to sizes of the order or slightly larger than the wavelength of the incident light.  Therefore, the solution of the problem remains unclear.

 Those are examples that illustrates that the retrieval of the physical characteristics of cosmic dust grains from the observed scattered light is far from trivial.  If the dust cloud of interest consists of spherical particles we can perform computations without any restriction about the size or composition by means of the Lorenz-Mie theory (\citet{vandehulst1957}). However the scattering properties of irregular cosmic dust grains can differ dramatically from those of equivalent spheres.  Due to their complicated morphology in most of the cases computations for polydisperse cosmic dust grains have to be replaced by simplified models such as spheroids (\citet{misha1997}, \citet{kim1995}), or hollow spheres \citep{min2005}.  Experimental data should then be used to validate the computational results.  Computations for dust grains of arbitrary shapes are more complicated and limited to particles with sizes comparable to the wavelength \citep{mackowski2011}. Thus, controlled laboratory experiments of light scattering by cosmic dust grains covering different size ranges, shapes and  compositions remain an indispensable tool to study the scattering behavior of irregular dust grains.  

In this work we present the measured phase functions for three different types of millimeter-sized cosmic dust  grains analogues. This size range is still poorly studied due to the limitations of the numerical codes and technical difficulties related to the experiments. The paper is organized as follows: in Section~\ref{sec:theory} we summarize the main concepts of light scattering providing a description of the experimental apparatus in Section~\ref{sec:setup}.  Test measurements regarding the reliability of the experimental data are presented in Section~~\ref{sec:test}.  In Sections~\ref{sec:samples} and \ref{sec:results}, we describe the physical characteristics of the dust grains and experimental phase functions, respectively. Finally, we summarize our results in Section~\ref{sec:conclusions}.  

\section{The scattering matrix formalism}\label{sec:theory}

The flux vector of the light scattered by one particle in a particular orientation, p,  is related to the flux vector of the incident beam, ${\pi \bf\Phi_{0}}$ by means of the scattering matrix, ${\bf F^{p}}$, as follows \citep{hovenier2004}:

\begin{equation}
  {\bf\Phi_{det}} = \frac{\lambda^{2}}{4\pi^{2}D^{2}} {\bf F^{p}}{\bf \Phi_{0}},
\end{equation}

\noindent where  $\lambda$ is the wavelength, and $D$ is the distance between the particle and the detector. Here, {$\pi \bf \Phi_{det}$} is the flux vector at the detector and ${\bf F^{p}}$ is the scattering matrix of the particle in a particular orientation. All elements of ${\bf F^{p}}$ are dimensionless and depend on the physical properties of the particle (size, shape, porosity, surface roughness, and refractive index), and the direction of scattering, i.e., the direction from the particle to the detector. The direction of scattering is defined by the scattering angle, $\theta$, the angle between the directions of propagation of the
incident and the scattered beams (0$\leq \theta\leq \pi$), and an azimuth angle, $\phi$, that ranges from 0 to 2$\pi$.    
In general {\bf $F^{p}$} contains 16 non vanishing elements:

\begin{equation}
{\bf F^{p}}= \left(  \begin{array}{c c c c}
                    F^{p}_{11}& F^{p}_{12} & F^{p}_{13}  & F^{p}_{14}  \\
                    F^{p}_{21}& F^{p}_{22} & F^{p}_{23} & F^{p}_{24} \\
                   F^{p}_{31}&F^{p}_{32}& F^{p}_{33}  & F^{p}_{34} \\
                   F^{p}_{41} &  F^{p}_{42} & F^{p}_{43}  & F^{p}_{44}
                   \end{array}\right).
\end{equation}

In the case of a particle in random orientation, all scattering planes are equivalent. Thus, the scattering direction is fully described by the scattering angle $\theta$. 
Further, for an homogeneous sphere the scattering matrix has only four independent elements that are not identically equal to zero, i.e. it has the form:  

\begin{equation}
{\bf F^{p}}= \left(  \begin{array}{c c c c}
                    F^{p}_{11} & F^{p}_{12} & 0 & 0  \\
                    F^{p}_{12} & F^{p}_{11} & 0 & 0 \\
                   0 & 0 & F^{p}_{33}  & F^{p}_{34} \\
                   0 & 0 & -F^{p}_{34}  & F^{p}_{33}
                   \end{array}\right).
\end{equation}

For unpolarized incident light, the first element of the scattering matrix, $F^{p}_{11}(\theta)$, is proportional to 
the flux of the scattered light and is  called the phase function
or the scattering function.

\section{Experimental Apparatus} \label{sec:setup}

\begin{figure}
\gridline{\fig{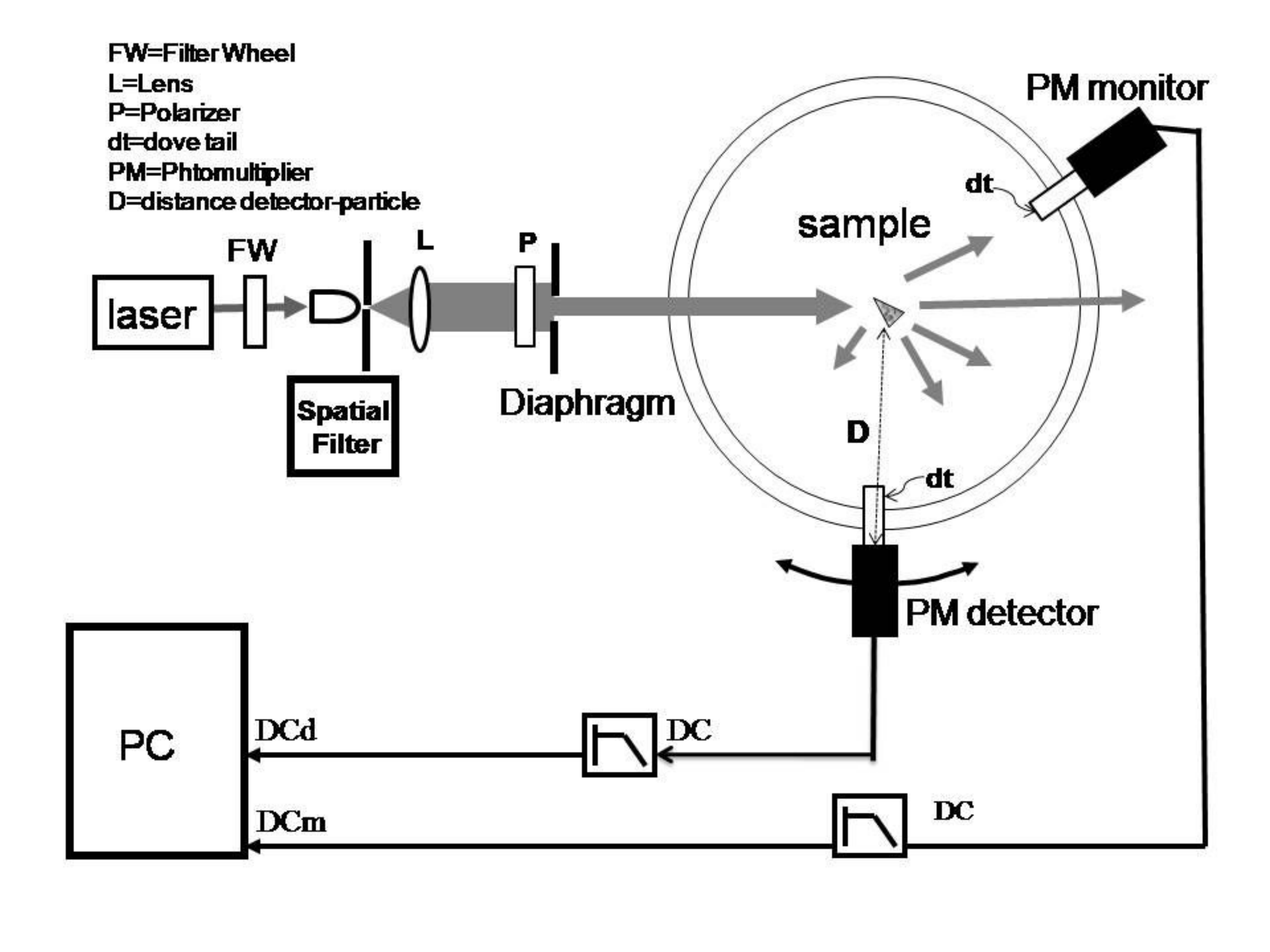}{0.45\textwidth}{(a)}
          \fig{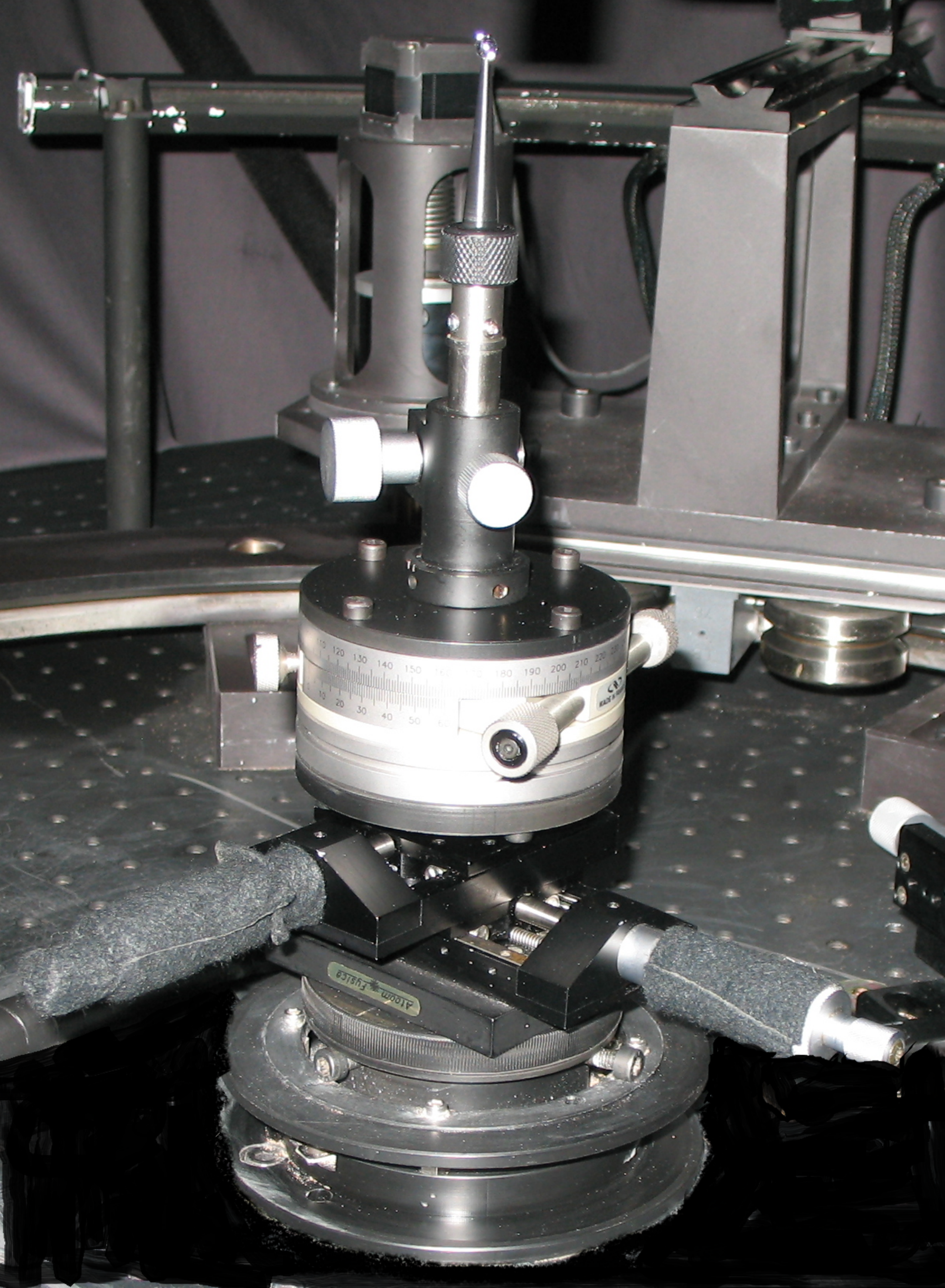}{0.25\textwidth}{(b)}
}
\caption{(a): Schematic overview of the experimental light scattering apparatus as seen from above. (b): Photograph of the flat black holder mounted on the x-y rotating table. The N-BK7 glass sphere is located on its conical tip.}
\label{fig:setup}
\end{figure}

The light scattering measurements have been performed at the IAA COsmic DUst LABoratory (\citet{munoz2010}, \citet{munoz2011}). 
In this work, the experimental apparatus has been adapted to measure the angular dependence of the flux scattered by a single particle with a size much larger than the wavelength of the incident light.
A schematic overview of the experimental apparatus is presented in Figure~\ref{fig:setup}, (a). We use a linearly polarized continuous-wave tunable Argon-Krypton laser tuned at 520 nm.  A spatial filter has been used to avoid spatial intensity variations in the laser beam. In this way we assure a homogeneous illumination over the entire particle. The homogeneous beam is collimated by a lens  before passing  through a polarizer (P) oriented at 45 degrees. The polarized beam is scattered by the particle of interest. This is located on a 2mm conical-tip flat black holder mounted on a x-y rotating table (Figure~\ref{fig:setup}, (b)). 
A filter wheel (FW) equipped with grey filters of different density is located between the laser and spatial filter. It is operated from the computer so that the flux of the incident beam can be scaled to its most appropriate value for each scattering angle. The unscattered part of the incident beam is absorbed by a beam-stop. Moreover, we use a diaphragm behind the polarizer (P) to control the width of the beam so that  only the particle of interest is illuminated and not the holder.  The scattered light is detected by a
photomultiplier tube (9828A Electron tubes\copyright), the detector.  Another photomultiplier tube (the monitor) is located at a fixed position and is used to correct for fluctuations in the laser beam. Both photomultipliers are positioned on a ring with an outer diameter of one meter. Detector and Monitor are mounted on dove tails (dt) so that they can be moved forward and backward. In this case the detector is located at a distance, D, of 62 cm to the particle. The detector moves along the ring in steps of 5 degrees, 1 degree or even smaller if a higher angular resolution is required, covering a scattering angle range from  3 degrees (nearly forward scattering) to 170 degrees (nearly backward scattering).

\section{Test Measurements}
\label{sec:test}
The reliability of the measurements is tested by comparing the measured phase function of two calibration spheres to results of Lorenz-Mie calculations for the corresponding size and refractive index. 
Physical properties of the N-BK7 glass and Sapphire spheres (Edmund Optics) are presented in Table~1. The size of the calibration spheres has been chosen similar to that of our particles of interest. 

\begin{table}[h!]
\label{tab:properties}
\renewcommand{\thetable}{\arabic{table}}
\centering
\caption{Properties of the calibration spheres and cosmic dust grains.}
\begin{tabular}{lllc}
\hline
Composition & diameter (mm) & m=n+ki (520 nm)& \\
\hline
N-BK7 & 5.0   & 1.5168+9E-9i  &  Edmund Optics Catalog \\
Sapphire &  5.0  &  1.77+0i   & Edmund  Optics Catalog \\
Enstatite & 6.4\footnotemark[1] & 1.58+2E-05i& \citet{dorschner1995}\\
Quartz & 7.8\footnotemark[1] & 1.54+0i & \citet{klein1993} \\
Etna & 7.0\footnotemark[1] & 1.59+0.01i & \citet{ball2015} \\  
\hline
\footnotetext{diameter of the volume-equivalent sphere.}
\end{tabular}
\end{table}

\begin{figure}
\gridline{\fig{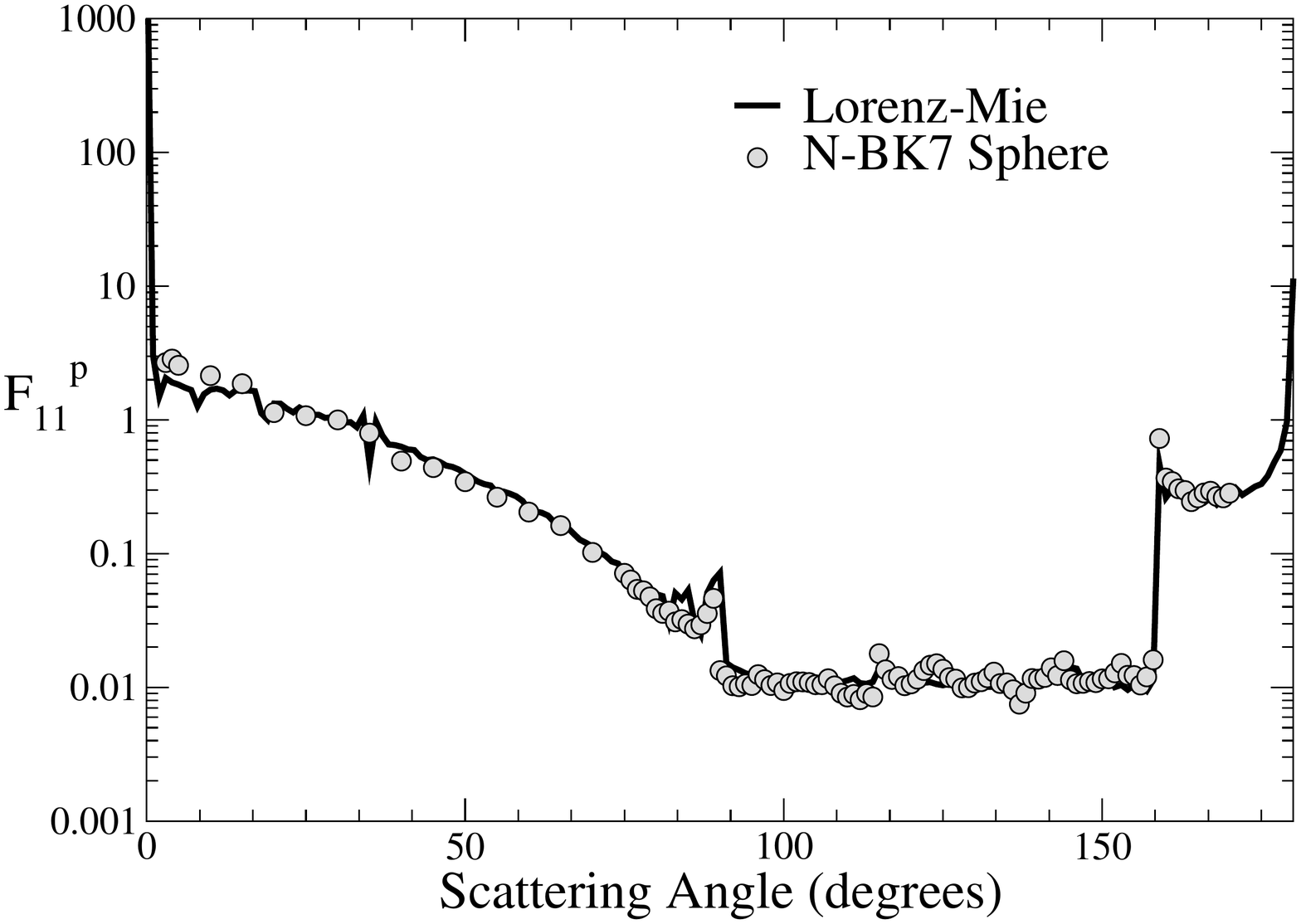}{0.45\textwidth}{(a)}
          \fig{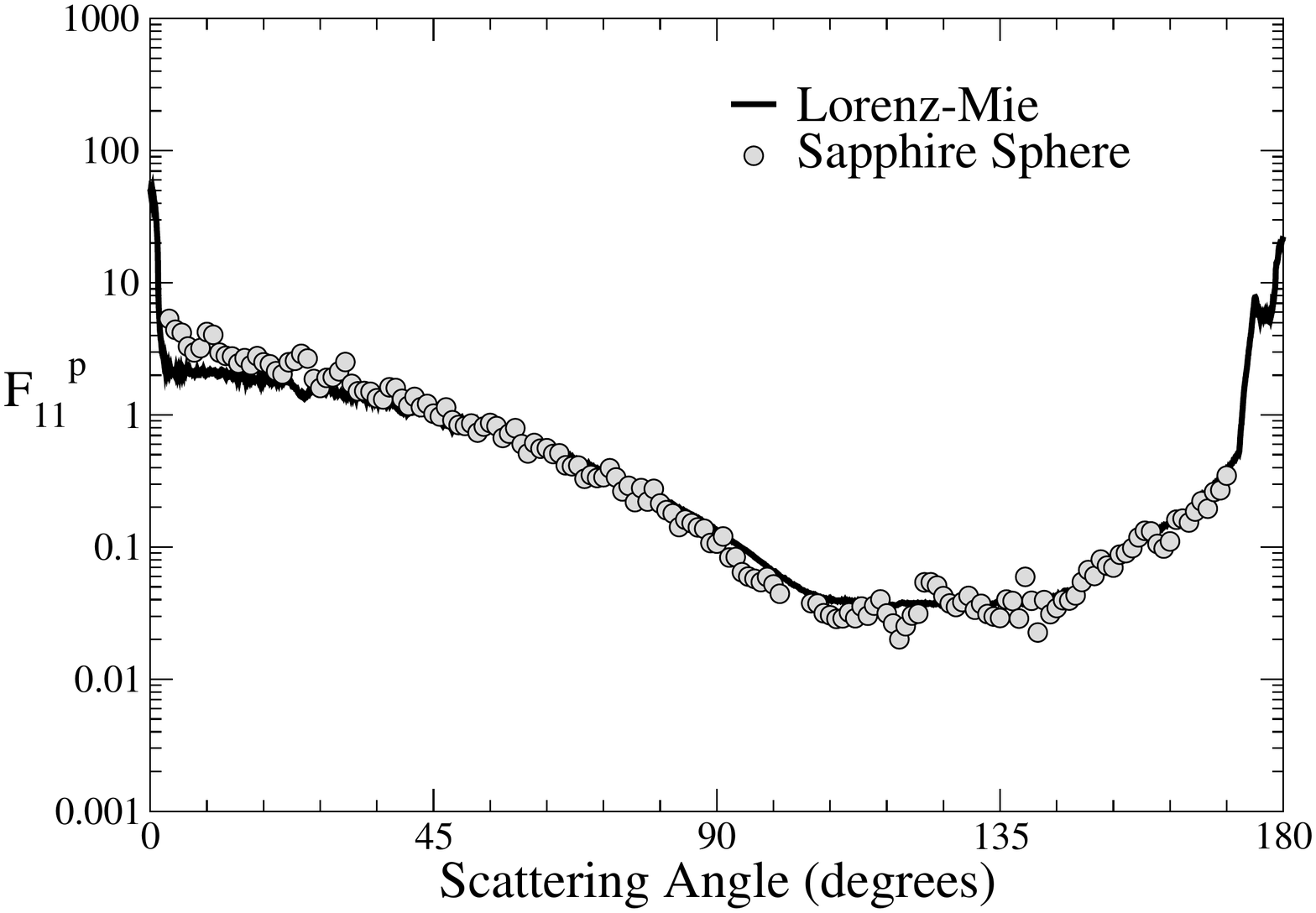}{0.45\textwidth}{(b)}
}
\caption{Comparison of phase functions based on the measured data and the Lorenz-Mie computations (solid black) for the N-BK7 (a) and Sapphire (b) calibration spheres (gray circles).}
\label{fig:spheres}
\end{figure}

In Figures \ref{fig:spheres} (a) and (b), we present the measured and calculated phase functions as functions of the scattering angle for the N-BK7 and Sapphire calibration spheres, respectively.  The measured and calculated 
$F_{11}^{p}(\theta)$ are plotted on a logarithmic scale and normalized to 1 at 30 degrees. 
During the test measurements the detector is moved along the ring in steps of 1 degree. 
The plotted values corresponding to Lorenz-Mie calculations are averaged over $\pm$ 0.25 degrees according to the angular resolution of the experimental setup. As shown in Figures~\ref{fig:spheres} (a) and (b), the measured phase functions show an excellent agreement with the Lorenz-Mie computations over the entire angle range. Small differences might be caused by small inhomogeneities in the calibration sphere. Moreover, the measured results are strongly dependent on the exact position of the calibration sphere at the center of the measuring ring.



\section{Dust grains}
\label{sec:samples}

In this work we study three types of mm-sized grains consisting of enstatite, quartz, and volcanic material from Mount Etna, respectively.  The three grains  have been chosen so that they would have similar sizes but different absorbing properties.  Mount Etna is a quite dark material whereas Enstatite and Quartz present a nearly zero imaginary part of the refractive index at the studied wavelength (527 nm).  In this way we can study how absorption affects the measured phase functions. Their physical properties are summarized in Table~1. Moreover, all dust grains present similar sizes as the BK7 and Sapphire spheres used to test the experimental apparatus.  In this way we assure a homogeneous illumination over the entire grain.

\begin{figure*}[ht!]
\gridline{\fig{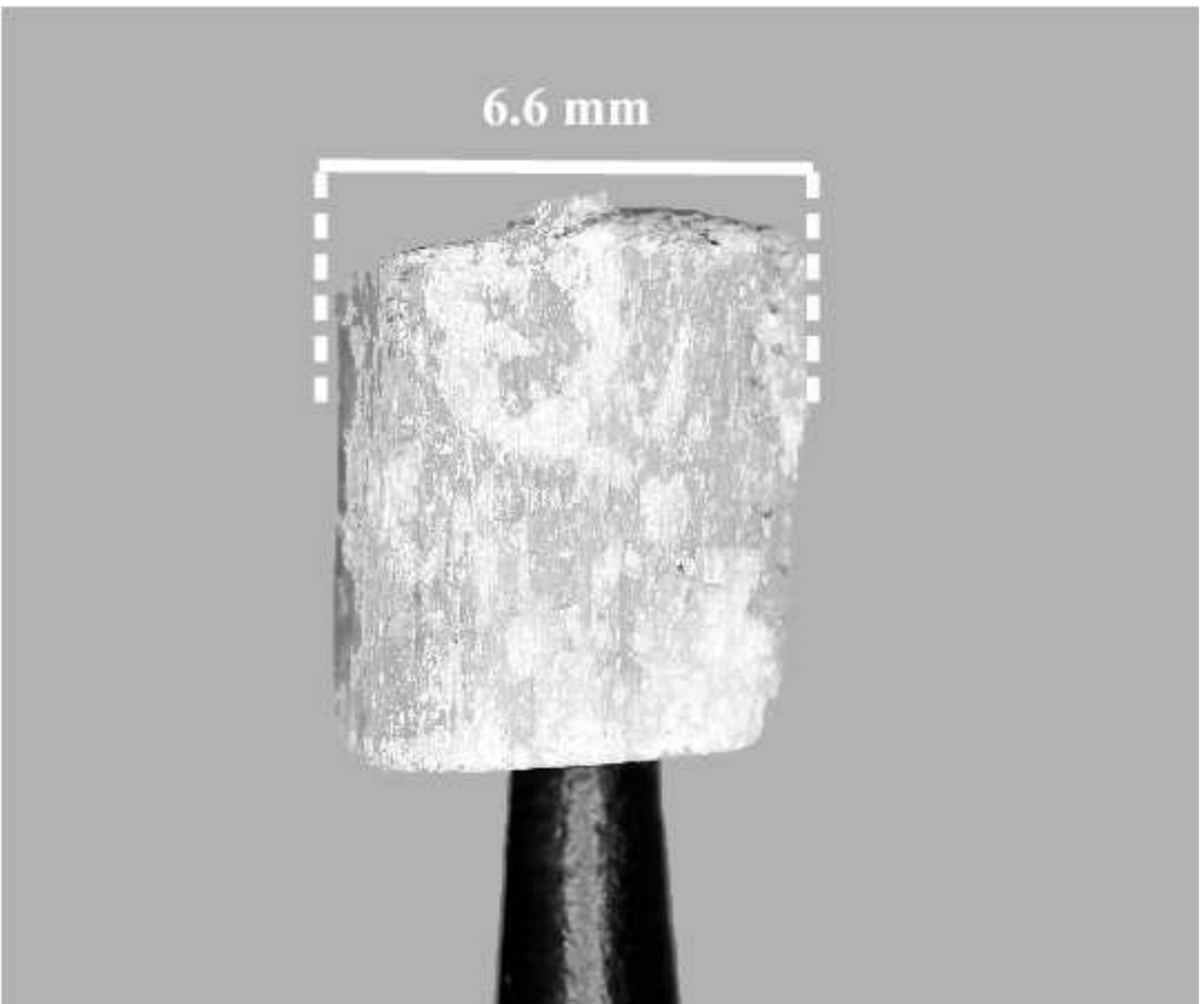}{0.3\textwidth}{(a)}
          \fig{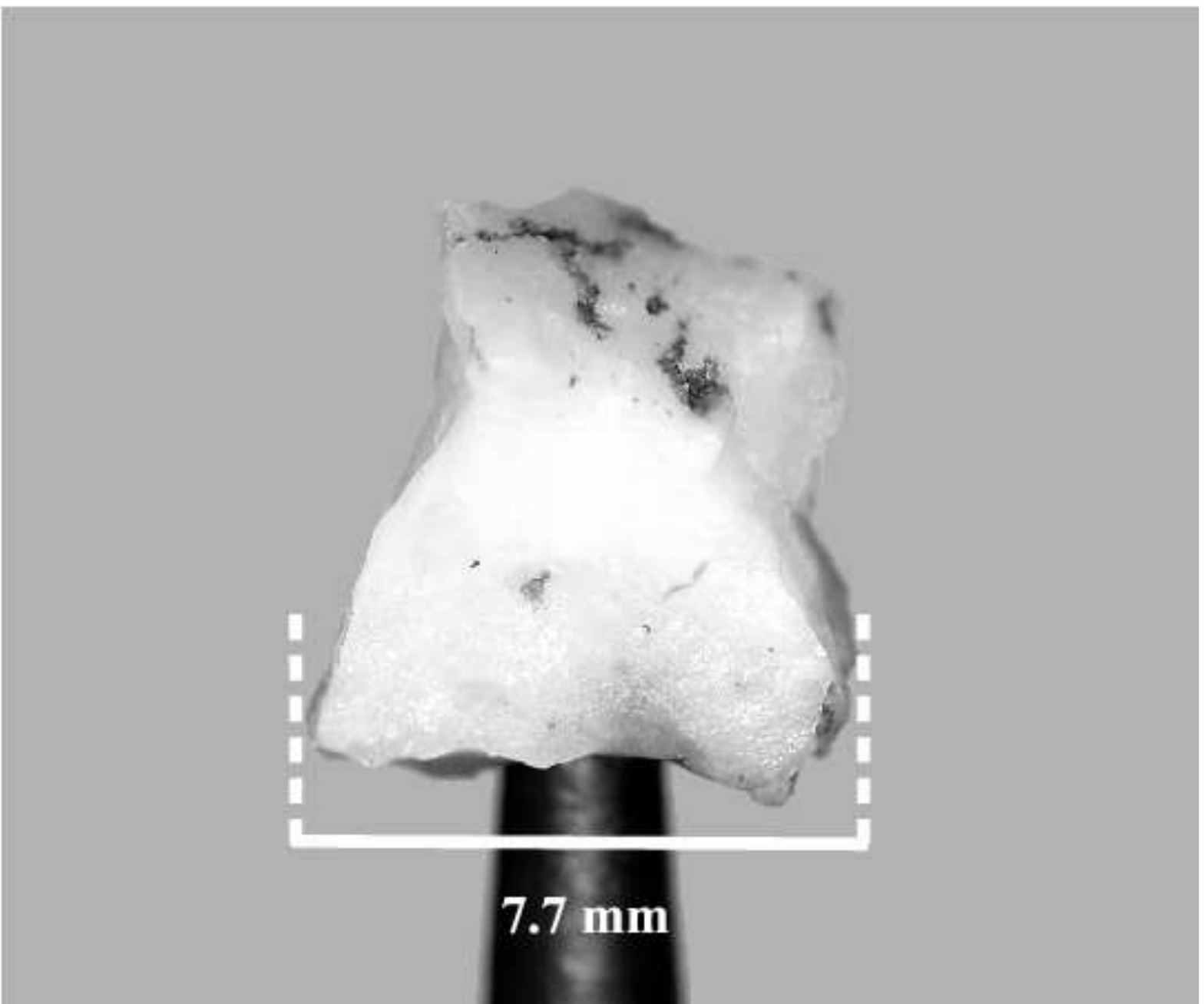}{0.3\textwidth}{(b)}
          \fig{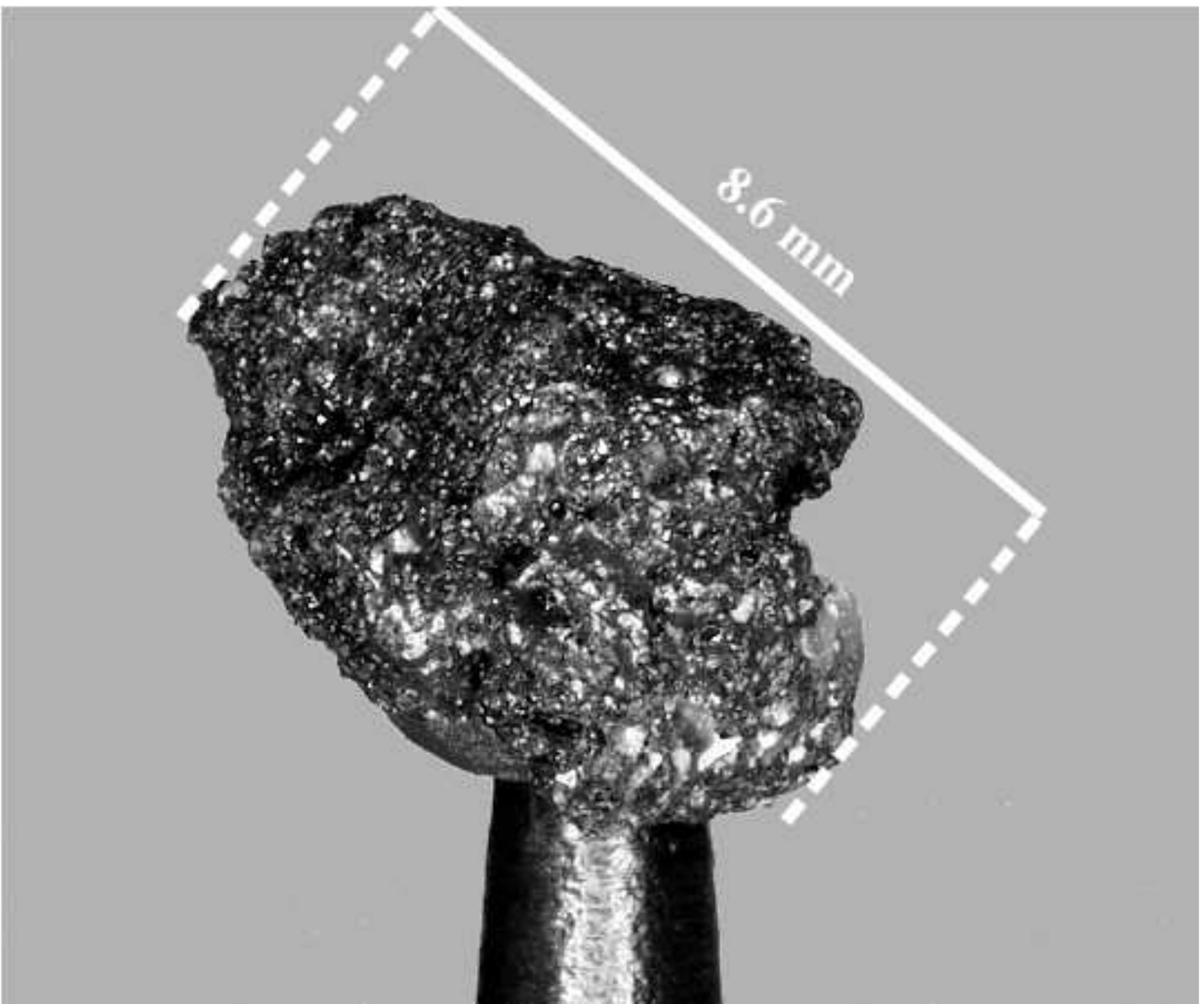}{0.3\textwidth}{(c)}
          }
\caption{Optical images of Enstatite (a), Quartz (b) and 
Etna (c) grains. \label{fig:camara}}
\end{figure*}

\begin{figure*}
\gridline{\fig{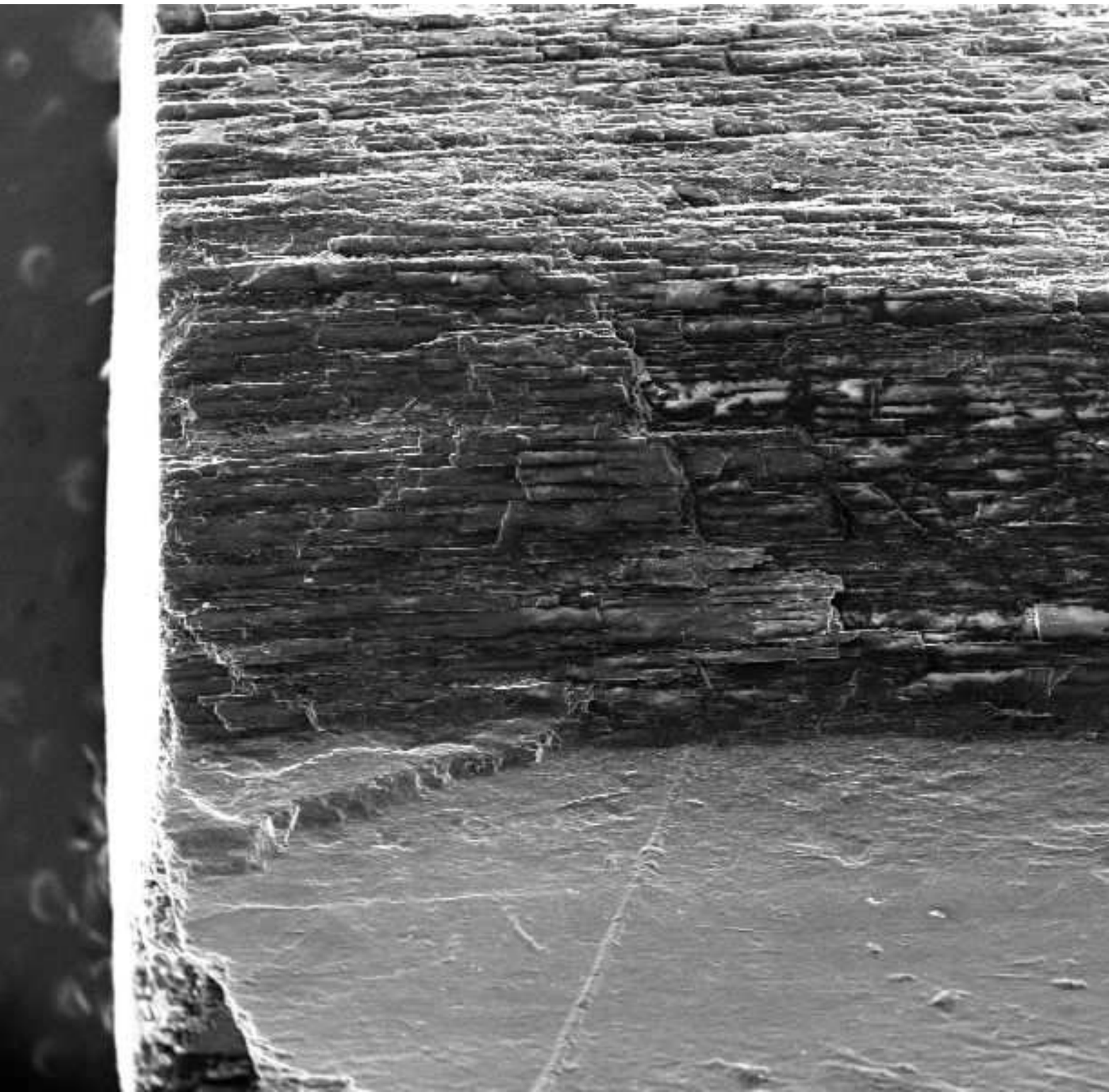}{0.3\textwidth}{(a)}
          \fig{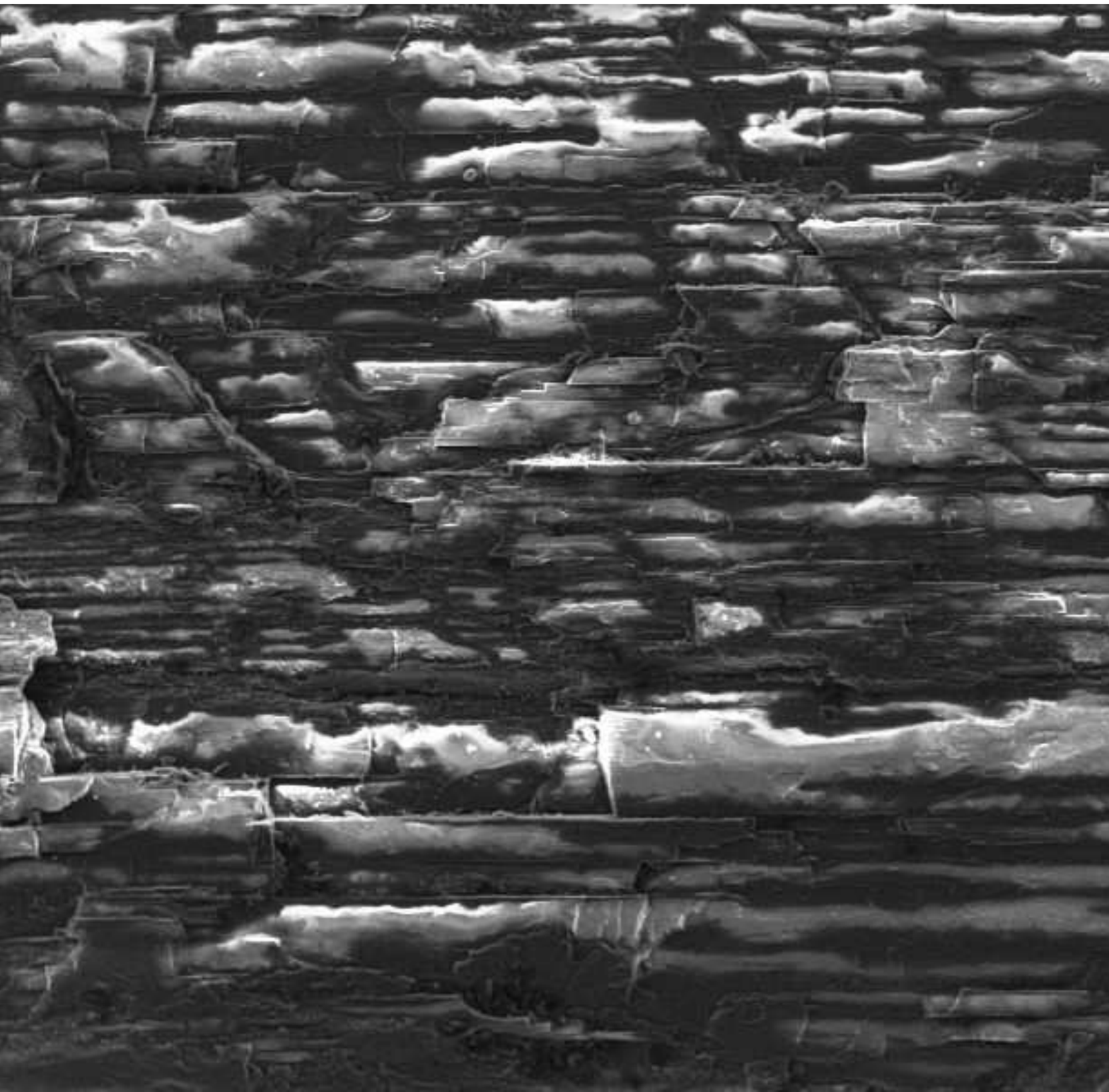}{0.3\textwidth}{(b)}
          \fig{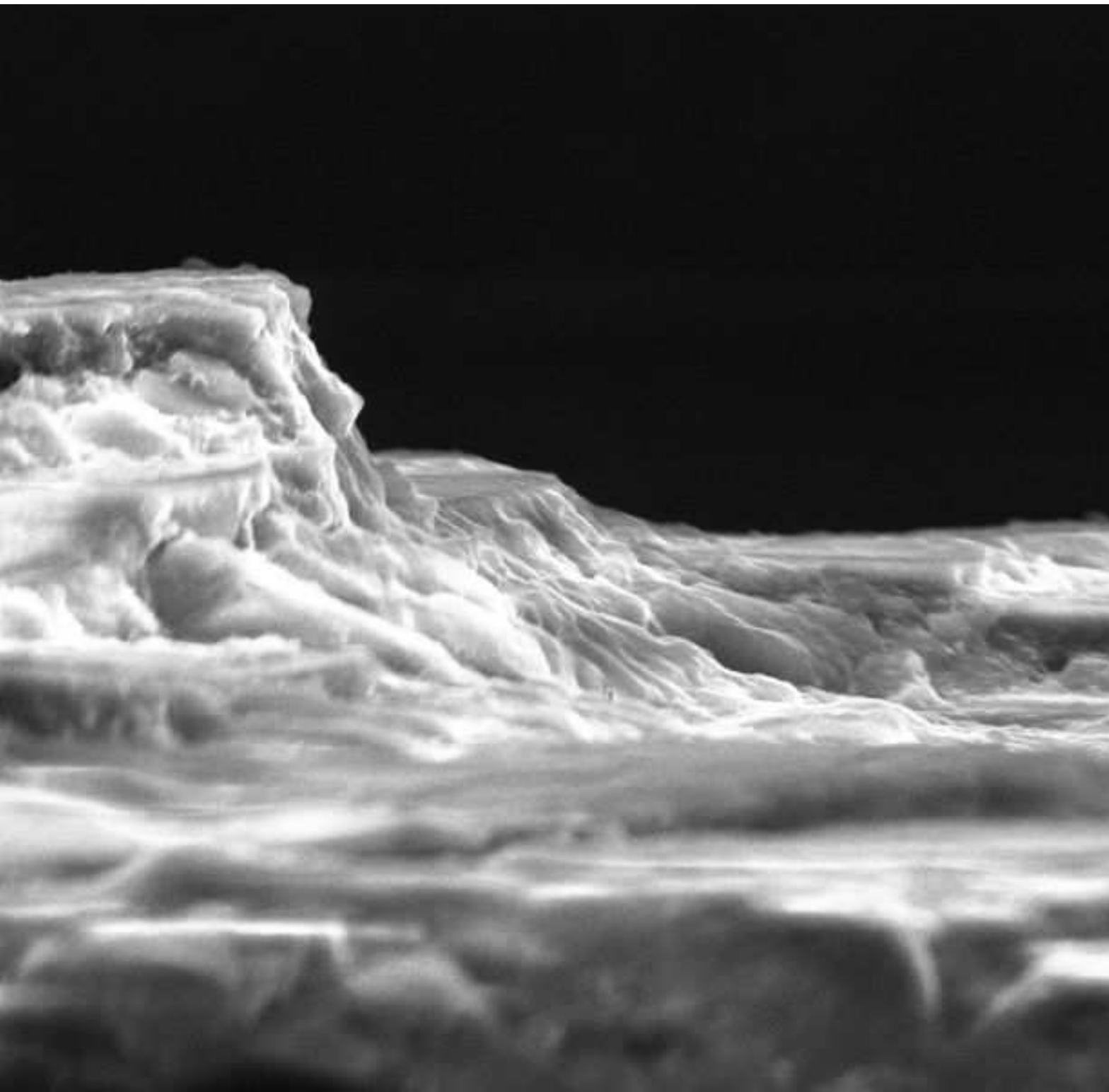}{0.3\textwidth}{(c)}
          }
\caption{Scanning Electronic Microscope images of Enstatite. White bars denote 500 $\mu$m (a), 100 $\mu$m (b), and 10 $\mu$m (c), respectively. \label{fig:ME_Ensta}}
\end{figure*}

\begin{figure*}
\gridline{\fig{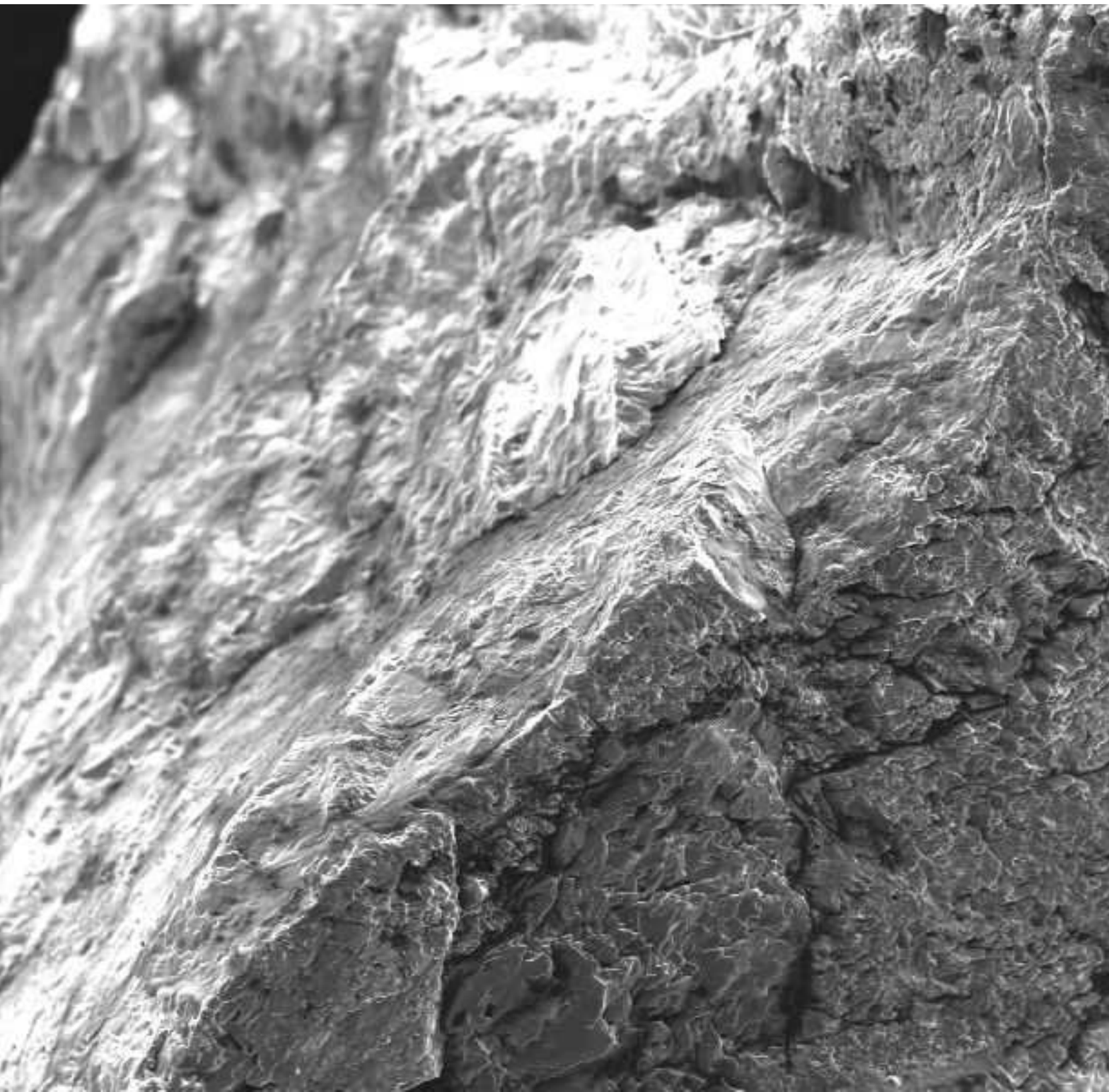}{0.3\textwidth}{(a)}
          \fig{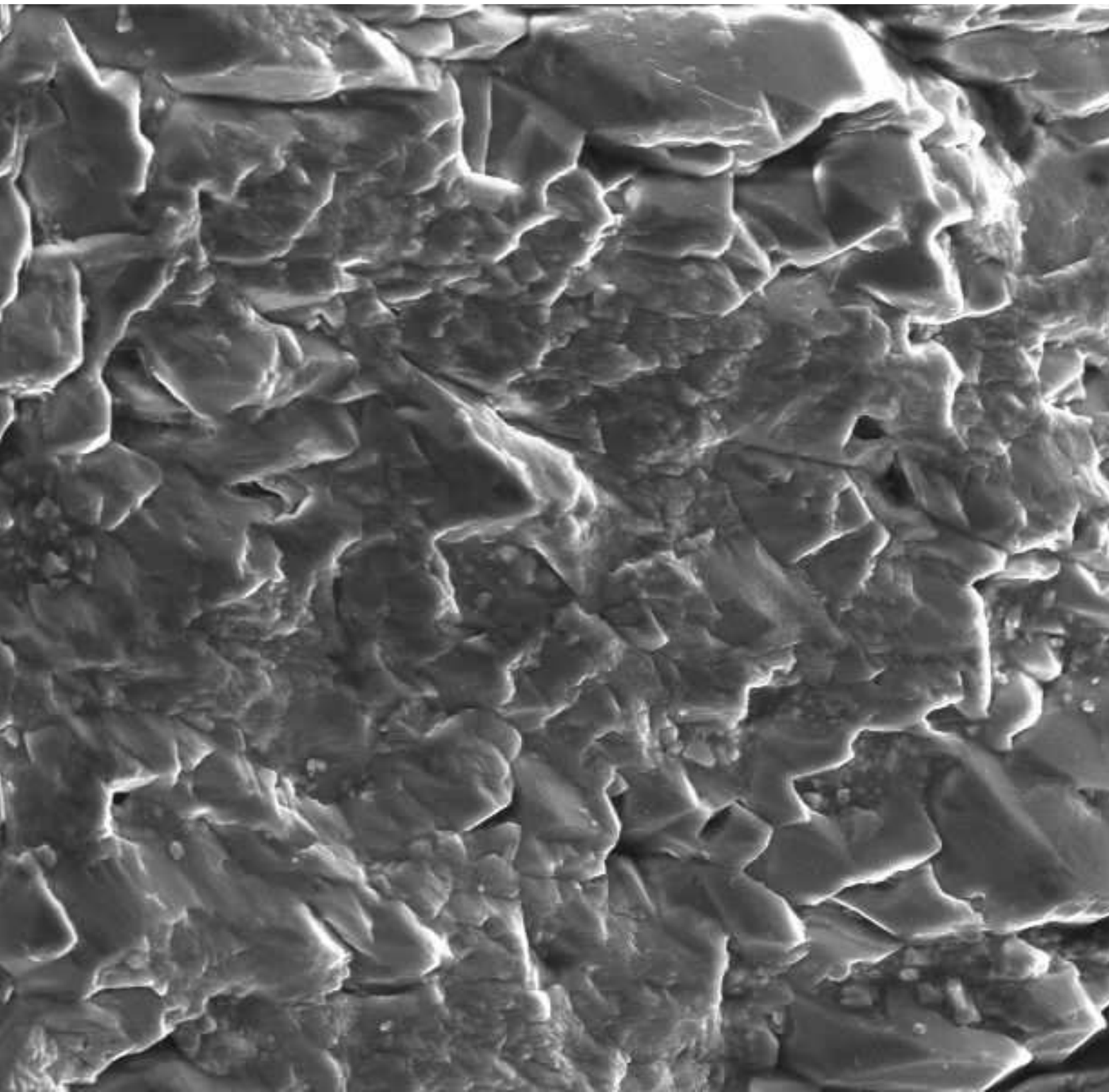}{0.3\textwidth}{(b)}
          \fig{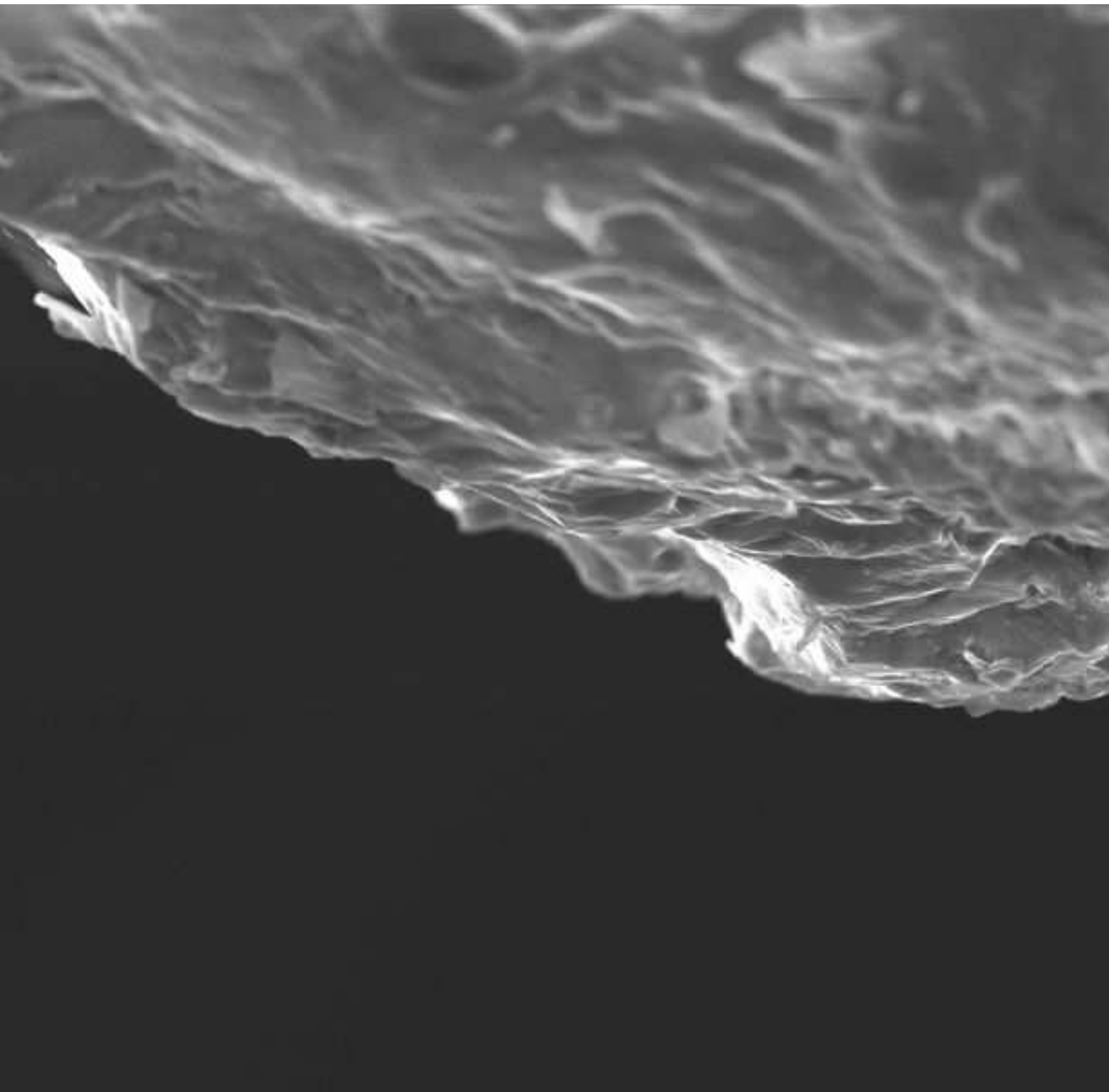}{0.3\textwidth}{(c)}
          }
\caption{Scanning Electronic Microscope images of Quartz. White bars denote 
500 $\mu$m (a), and 50 $\mu$m (b) and (c), respectively.  
\label{fig:ME_Quartz}}
\end{figure*}

\begin{figure*}
\gridline{\fig{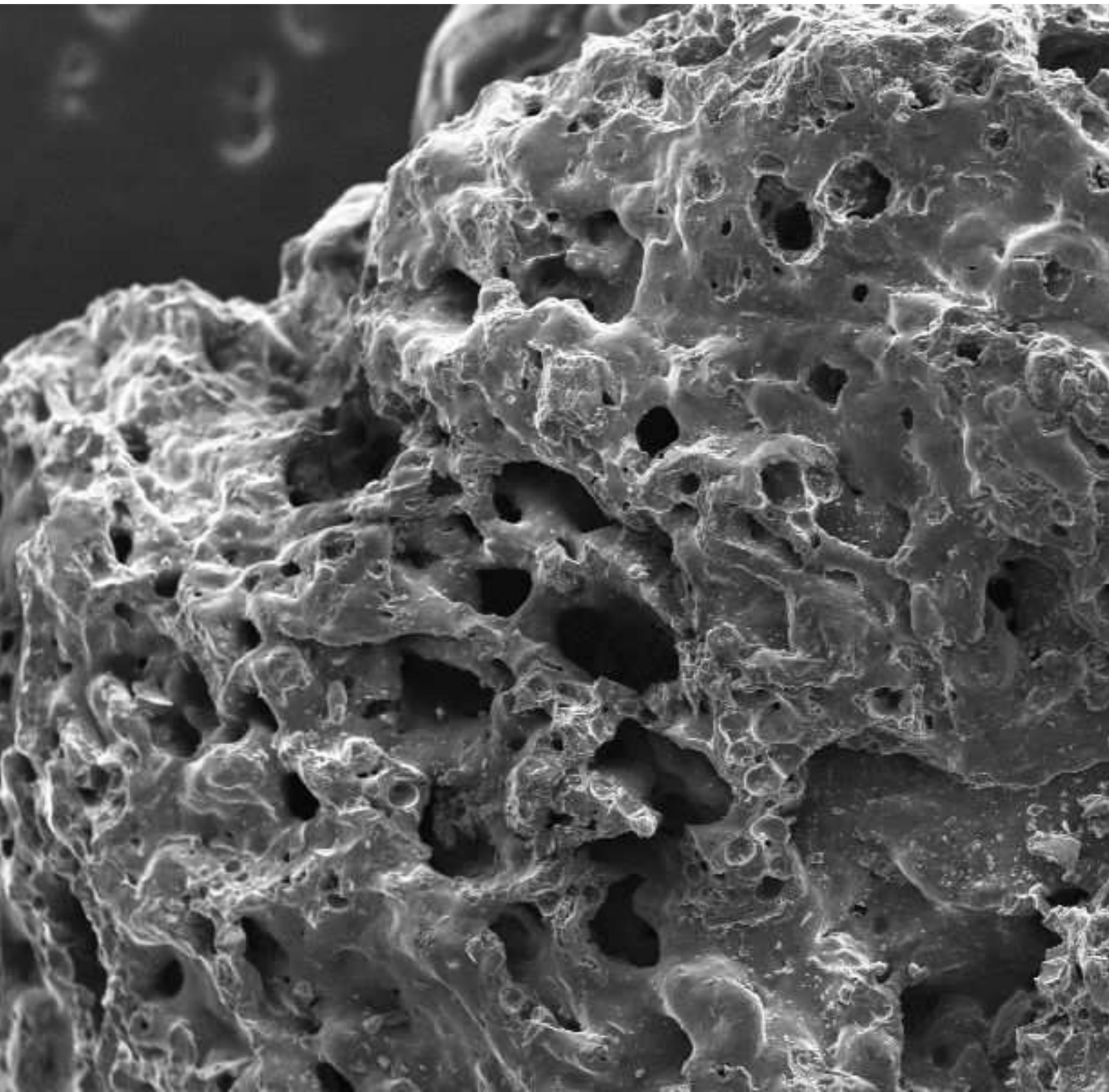}{0.3\textwidth}{(a)}
          \fig{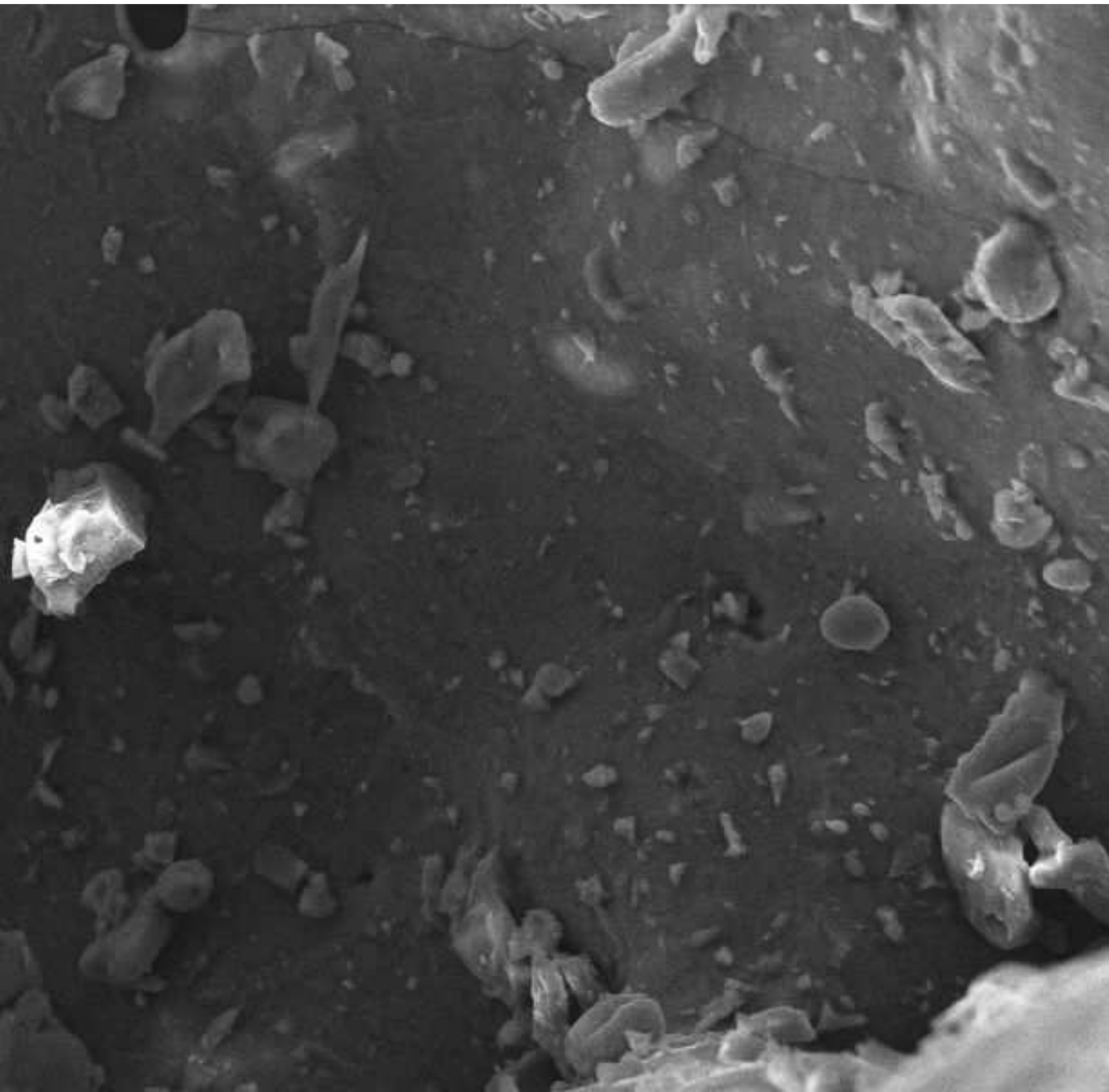}{0.3\textwidth}{(b)}
          \fig{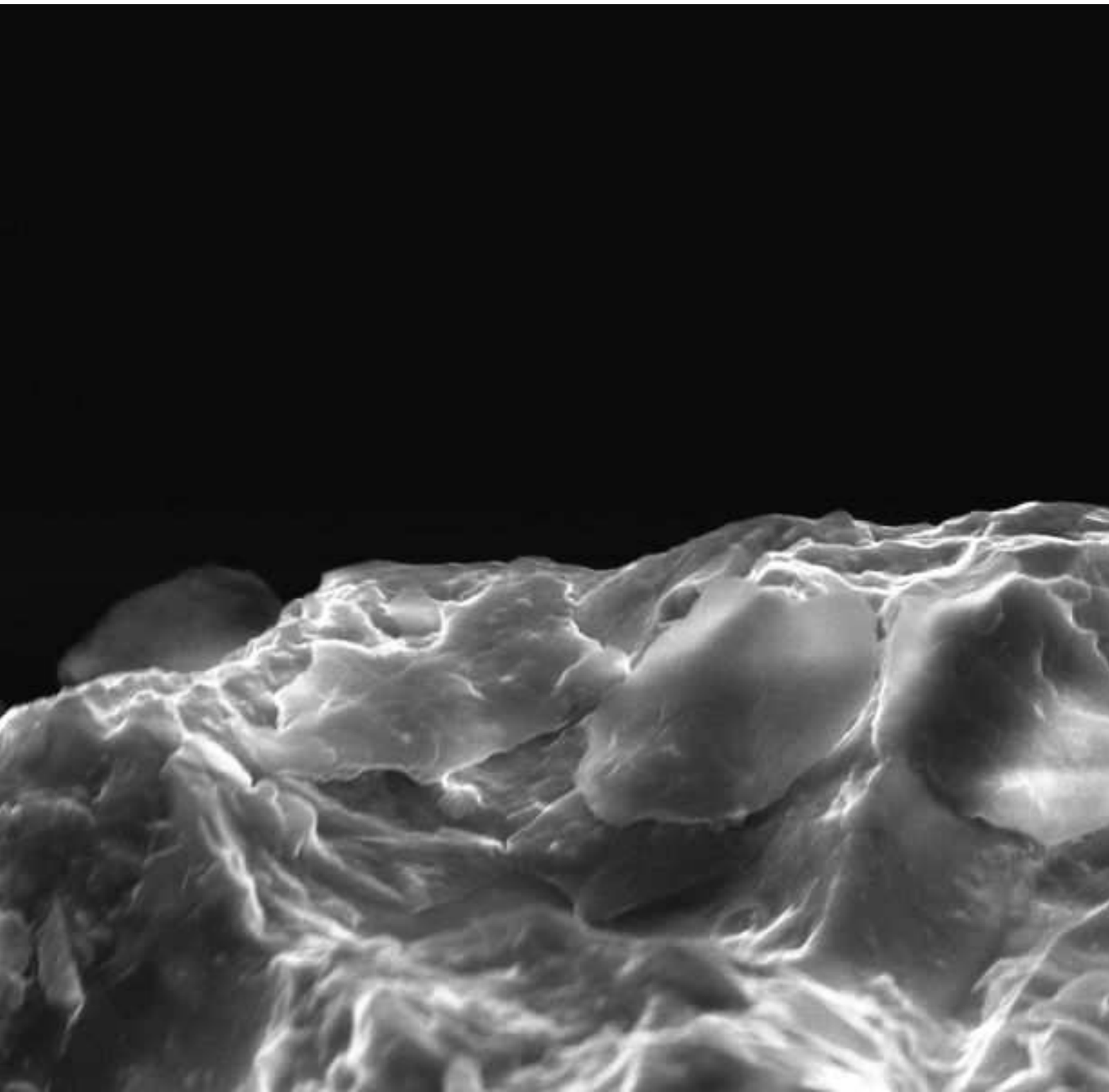}{0.3\textwidth}{(c)}
          }
\caption{Scanning Electronic Microscope images of the Etna grain. White bars denote 
500 $\mu$m (a), 50 $\mu$m (b), and 10 $\mu$m (c), respectively.  
\label{fig:ME_Etna}}
\end{figure*}

In Figure~\ref{fig:camara} we present optical images of the three dust grains.
It is interesting to note that even in the optical images we can distinguish surface roughness.  This is in particular the case of the Etna grain that presents the typical structure of vesicular volcanic grains composed of porous material and cratery surfaces \citep{riley2003}. The vesicular structure is produced by gas bubbles that scape when the volcanic melt is cooled to glass. To estimate the scale of the surface roughness of our dust grains in Figures~\ref{fig:ME_Ensta}-\ref{fig:ME_Etna} we present Scanning Electronic Microscope images. As shown the three studied grains present not only intrinsic surface roughness with sizes of the order or smaller than the wavelength of the incident light, but also micron-sized grains that might affect their optical properties  (Figures~\ref{fig:ME_Ensta}-\ref{fig:ME_Etna}, top right panels).

\section{Results and discussion} \label{sec:results}

\begin{figure}
\begin{center}
\rotatebox{0}{
\resizebox{0.8\textwidth}{!}
{
\includegraphics{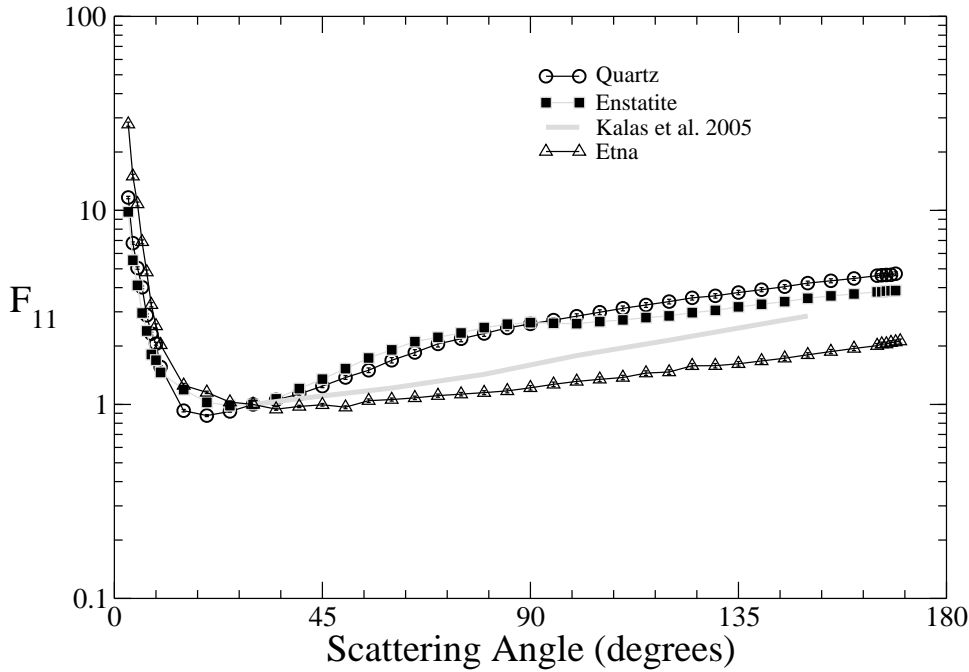}
}}
\caption{Experimental phase functions at 520 nm for Quartz (filled circles), Enstatite (filled squares), and Etna (triangles). The observed phase function of the Fomalhaut disk grains from \citet{kalas2005} is also shown. All phase functions are normalized to unity at 30 degrees.  Errors are indicated by error bars or are within the symbols. \label{fig:stones}}
\end{center}

\end{figure}

In Figure~\ref{fig:stones} we present the measured phase functions for the Enstatite, Etna and Quartz dust grains at 520 nm. The measurements are presented together with the observed phase function of the Fomalhaut dust grains obtained with the Henyey-Greenstein parametrization retrieved by \cite{kalas2005}. The Henyey-Greenstein phase function is constrained to the observable range of the Fomalhaut system. All phase functions are normalized to unity at 30 degrees scattering angle and are presented on a logarithmic scale.

The procedure of the measurements is as follows: The detector is moved along the ring in steps of 1 or 5 degrees covering the scattering angle range from 3 to 170 degrees. During the measurements the particles are located on the rotating conical holder shown in Figure~\ref{fig:setup}, (b). A rigorous 3D orientation average could be obtained by a sufficient number of Euler rotations: i. Rotation  around the vertical axis, $\alpha$, ii. Rotation around the direction of the laser beam, $\beta$, and  iii. Rotation around the axis perpendicular to the base of the particle, $\gamma$. For an optimal performance, $\alpha$ and $\gamma$ should be uniformly distributed while $\beta$ ought to follow a distribution proportional to sin($\beta$) \citep{misha2017}. Such a procedure would require a holder with two degrees of freedom that would significantly complicate the performance of the measurements. Instead, we have assumed a 1-axis orientation average as an approximate solution. This is equivalent to a 3D orientation average in the case that the revolution volume of the particle is symmetrical with regard to the scattering plane. This condition is fulfilled up to a high degree by our cosmic grains as shown in Figure~\ref{fig:camara}. Thus, to simulate random orientation the plotted figures, $F_{11}(\theta)$, are the results of averaging over 36 $F_{11}^{p}(\theta)$, corresponding to 36 different orientations. Starting with a given orientation, a particle is measured after each of 36 successive rotations of 10 degrees. We assume that the number of orientations is sufficient when adding more orientations of the particle on the vertical axis (perpendicular to the direction of the detector) does not affect the final result.  Special tests are performed to verify that other directions are irrelevant for such irregular grains. In those tests the position of the grain on the holder is rotated 90 degrees about a horizontal plane. Starting from a given position, the grain was measured after each of two 90 degrees rotation about the vertical axis.  
 It is verified that adding those extra orientations does not affect the final result. In conclusion, taking into account the 
orientation over the horizontal axis plays no significant role in our experimental results. That is a good indication that the measurements can be considered in random orientation even though strictly speaking the phase function has not been averaged over all possible orientations.

In all three measured phase functions we can distinguish two well defined regions. Firstly, soft forward peaks in the 3 to $\sim$20 degrees. Secondly, in the scattering angle range from  $\sim$20 to 170 degrees, the three phase functions increase with the scattering angle. This increase is stronger in the case of the two non-absorbing grains, Enstatite and Quartz. For the sizes of our dust grains the ray optics approximation can be used. It is based on  the assumption that the incident plane wave can be represented as a collection of independent parallel rays. The rays hitting the particle result in two phenomena i) diffraction, that is constrained into a narrow intensive lobe around the exact forward direction (0 degrees) and ii) reflection and refraction that contribute to the total scattering by the particle \citep{vandehulst1957}.   

The angular width of the diffraction peak for the size of our dust grains and distance to the detector, D,  is of the order of $\pm$0.8 degrees around the exact forward direction that is beyond the measurable angle range of our experiment. Therefore, diffraction by the large grains cannot be responsible for the measured forward peaks.  

It is known that wavelength-scale surface roughness can significantly affect the scattering properties of dust grains \citep{munoz2007}, \citep{nousiainen2011}, \citep{lindqvist2011}, \citep{escobar2017}.  As presented in Figures~\ref{fig:ME_Ensta}-\ref{fig:ME_Etna} our particles are covered not only by various types of surface structures and cavities but also by small particles.  Therefore the measured forward peaks in the 3 to $\sim$ 20 degrees scattering angle range might be due to scattering by the wavelength-scale surface roughness and surface micron-sized particles of our grains.

As mentioned, at side- and back-scattering regions all measured $F_{11}(\theta)$ increase with the scattering angle. In the case of the Etna grain, that presents a high imaginary part of the refractive index the transmitted part can be ignored and therefore the measured phase function in the $\sim$20-170 degrees region might be due to reflexion on the surface of the particle. That is not the case for the enstatite and quartz grains in which refracted light, after another refraction, may emerge from the particle, contributing to the measured intensity at side- and back-scattering regions producing the measured higher slope of $F_{11}(\theta)$ in the mentioned region.   

As shown in Figure~\ref{fig:stones}, the empirical phase function of the grains orbiting Fomalhaut lies within the domains occupied by the measured phase functions for the mm-sized non-absorbing and highly absorbing dust grains, respectively. That seems to indicate that the Fomalhaut dust ring could be dominated by very large grains. 

\begin{figure}
\begin{center}
\rotatebox{0}{
\resizebox{0.8\textwidth}{!}
{
\includegraphics{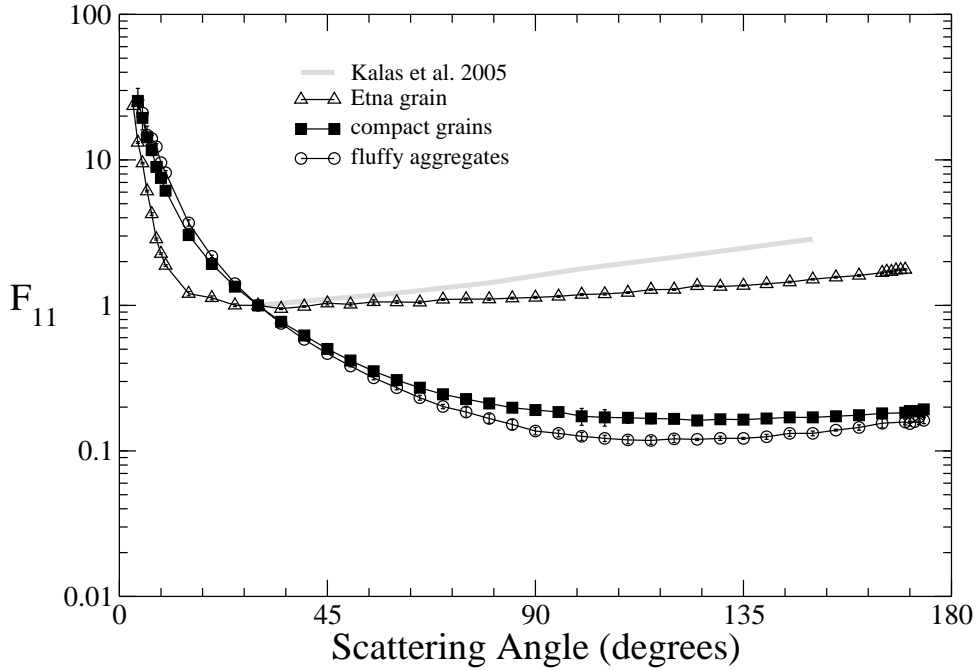}
}}
\caption{Experimental phase functions for the Etna dust grain (r=3.5mm) triangles, compact particles (r$_{eff}$=124.75 $\mu$m) \citep{munoz2007}, and micron-sized fluffy aggregates (circles) \citep{volten2006}. The observed phase function of the Fomalhaut disk grains from \citet{kalas2005} is also shown. All phase functions are  normalized to unity at 30 degrees.  Errors are indicated by error bars or are within the symbols.}\label{fig:stonesdust}
\end{center}

\end{figure}

Several model particles have been suggested to reproduce the HST optical and/or Herschel far-infrared images of the Fomalhaut system. Dust grains should be large enough so that the diffraction spike is narrowly forward peaked and therefore outside of the observable angle range.  The analysis of HST optical images performed by \citet{min2010} establishes a lower limit diameter of 100 $\mu$m for the grains in the Fomalhaut dust ring. Further studies including analysis of Herschel far-infrared images add another constraint to the dust grains populating the disk. They should simultaneously scatter light like large grains and absorb and emit like small grains.  That calls for  more sophisticated model particles such us fluffy aggregates \citep{acke2012} that in principle could fulfill both conditions.  In Figure~\ref{fig:stonesdust} we present the experimental phase functions for three different samples of cosmic dust analogues namely, a size distribution of silicate-type compact particles \citep{munoz2007}, a size distribution of fluffy aggregates \citep{volten2007}, and the Etna grain presented in this work. The experimental data are presented together with the retrieved phase function for the Fomalhaut disk grains. 
The compact sample consists of a silicate dust particles collected in the Sahara desert (Libya). Its refractive index at the measurements wavelength (632.8 nm) is equal to $m=1.5+i0.0004$ similar to that found in enstatite, an iron-free pyroxene \citep{dorschner1995}. Its  effective radius, $r_{eff}$,  and variance, $v_{eff}$ are equal to 125 $\mu$m and 0.15, respectively. Therefore, the dust grains of this sample show sizes larger than the lower limit established by \citep{min2010}. Further details can be found in \cite{munoz2007}. 

The aggregate sample was produced in a condensation flow apparatus in an experiment intended to mimic the formation of circumstelar dust. The phase function presented in Figure~\ref{fig:stonesdust} corresponds to a magnesiosilica sample labeled as Aggregate1 in \citep{volten2007}.  The measurements are performed at 632.8 nm. The aggregate size is estimated to be of the order 20 $\mu$m with modal grain sizes ranging from 50 nm to 120 nm depending on its composition  i.e. it consists of aggregates with sizes similar to those estimated for the grains 
populating the HR 4796A dust ring. A detailed description of the sample is provided in \citep{volten2007}. The measured phase function for the compact and aggregate samples are freely available in the Amsterdam-Granada light scattering database (www.iaa.es/scattering) under request of citation of \citep{munoz2012} and the paper in which the data were published.

As shown in  Figure~\ref{fig:stonesdust}, both the compact and aggregate samples produce a narrow diffraction spike compatible with the HST Fomalhaut  images. However, in both cases the measured phase functions decrease at side-scattering angles showing a nearly flat dependence in the scattering angle range from 90 to 177 degrees.  This seems to be a general trend for irregular dust particles with sizes ranging from sub-micron up to hundred microns \citep{munoz2012}. That behavior does not agree with the observed slope of the phase function for Fomalhaut and HR 4796A grains. On the contrary, the phase function for the millimeter-sized Etna grain shows a nearly perfect fit to the observations. That seems to support the hypothesis that the Fomalhaut and HR 4796A dust rings could be dominated by  dust grains significantly larger than hundred microns. In the case of Fomalhaut, those large grains should present a fractal structure to mimic the observed far-infrared spectra.  However, it is not clear yet if such 
 large aggregates can reproduce the observed  phase function.  Unfortunately, computations for aggregates are limited to sizes significantly smaller than the size of the grains expected to be in protoplanetary and debris disks as the finding of the Rosetta mission indicate (e.g. \citep{fulle2015}, \citep{rotundi2015}, \citep{hilchenbach2016}, \citep{mannel2016}).  Even with ever-increasing  algorithms sophistication, light scattering computations for dust grains of arbitrary shapes are still limited to particles with sizes comparable to the wavelength \citep{mackowski2011}. Thus, further experimental phase functions of millimeter-sized aggregates consisting of micron-sized monomers are needed to know if such large aggregates could produce the observed slopes of the phase functions in the Fomalhaut and HR 4796A dust rings.

\section{Conclusions} \label{sec:conclusions}

We present the experimental phase function of three millimeter-sized cosmic dust analogues. The measurements are performed at 520 nm covering the scattering angle range from 3 to 170 degrees. The reliability of the experimental apparatus has been tested by comparison of the measured phase function of two calibration spheres from Edmund Optics with Lorenz-Mie computations for the corresponding size and refractive index.  The three studied grains consists of enstatite, quartz and volcanic material from Mount Etna. In all studied cases the measured phase functions show two well defined regions: i) a soft forward peak and ii) a continuous increase with the scattering angle at side- and back-scattering regions. That increase is stronger in the case of the non-absorbing grains, namely enstatite and quartz. 

Experimental data presented in this work indicate that the scattering in the disk around Fomalhaut and HR 4796A could be dominated by large irregular cosmic dust grains. In the case of Fomalhaut, the combination of this conclusion with that based on the analysis of the far-infrared spectra as reported by \citet{acke2012}, would drive us to a model of dust grains consisting of fractal aggregates with sizes significantly larger than the wavelength of the incident light. Such large particles are in agreement with last findings obtained for comet 67P/Churyumov-Gerasimenko, target of the ESA Rosetta mission (e.g. \citep {fulle2015}, \citep{rotundi2015}, \citep{hilchenbach2016}, \citep{mannel2016}).

Further experiments and computations with millimeter-sized fluffy aggregates will be needed to draw some conclusions about the fluffy/compact nature of such large dust grains. Polarization laboratory measurements and observations also appear to be a good diagnostic tool for retrieving the nature of such dust grains, since light is on average scattered more within compact particles decreasing the degree of linear polarization \citep{xing1997}. 
The measured phase functions will be freely available in digital form in the Amsterdam-Granada light scattering database (www.iaa.es/scattering) under request of citation of \citet{munoz2012} and this paper. 

\acknowledgments
 Comments of an anonymous referee on an earlier version of this paper are
gratefully acknowledge. We are indebted to Roc\'{\i}o M\'{a}rquez from the Scientific Instrumentation center of the University of Granada for providing the SEM images. This work been supported by the Plan Nacional de Astronom\'{\i}a  y Astrof\'{\i}sica contracts AYA2015-67152-R and AYA2015-71975-REDT.


\listofchanges

\end{document}